\documentclass{article}
\usepackage{spconf,amsmath,graphicx,hyperref}
\usepackage{bm}
\usepackage{cite,url}
\usepackage{amssymb}
\usepackage{color}
\usepackage{siunitx}
\usepackage{tabularx}
\usepackage{booktabs}
\usepackage{multirow}
\usepackage{tikz}
\usetikzlibrary{arrows.meta,positioning}

\newcommand{\RR}{I\!\!R}

\newcommand{\secref}[1]{Section~\ref{#1}}
\newcommand{\eqnref}[1]{\eqref{#1}}
\newcommand{\figref}[1]{Fig.~\ref{#1}}
\newcommand{\tabref}[1]{Table~\ref{#1}}

\newcommand{\bx}{\mathbf{x}}

\newcommand{\bB}{\mathbf{B}}
\newcommand{\bC}{\mathbf{C}}
\newcommand{\bA}{\mathbf{A}}

\newcommand{\bI}{\mathbf{I}}
\newcommand{\bu}{\mathbf{u}}

\title{Unsupervised Instantaneous Phase and Frequency Tracking \\ by Inverse Voice Synthesis}
\name{Chin-Yun Yu \qquad Gy{\"o}rgy Fazekas}
\address{Centre for Digital Music\\
Queen Mary University of London\\
Mile End Road, London E1 4NS, UK
}
\begin{document}
\ninept
\maketitle
\begin{abstract}
  Knowledge-driven neural vocoders struggle to learn reliable fundamental frequency end-to-end, because spectral objectives provide weak supervision of periodic structure and lack phase information.
  We address this with a source-filter model whose alias-free additive source makes the instantaneous phase of the glottal cycle explicit; differentiating it yields the instantaneous frequency, and thus $F_0$, without an external tracker.
  Waveform error supervises only the deterministic harmonic path, while a spectral loss covers the full signal.
  On M4Singer and LM-SSD, the reconstruction is phase-aligned, reaching a signal-to-reconstruction-error ratio of \SI{8.1}{\dB}, while neural baselines remain negative.
  However, GOLF, given an external $F_0$, still reaches lower spectral distortion.
  On LM-SSD, the recovered $F_0$ attains the highest overall accuracy of any method tested, including supervised neural pitch trackers applied off the shelf, and the glottal closure instants come within 0.53 points of REAPER's identification rate, without any $F_0$ label.
\end{abstract}

\section{Introduction}\label{sec:introduction}

The fundamental frequency ($F_0$) is the rate of vocal fold vibration.
It shapes intonation in speech and singing and is a key control in parametric voice synthesis~\cite{zen2009parametric,moriseWORLDVocoderBasedHighQuality2016}.
These systems work by analysis-by-synthesis: parameters are estimated from the input, then used to rebuild it with a source-filter model~\cite{markelLinearPredictionSpeech1976}.

Recent neural vocoders integrate parameter extraction with neural networks~\cite{fabbroSpeechSynthesisControl2020,liuNeuralHomomorphicVocoder2020,wu1DDSPbasedSingingVocoders2022}, but learning $F_0$ end-to-end is hard because magnitude-spectral losses drop phase and give noisy gradients for periodicity~\cite{turianSorryYourLoss2020}.
Prior systems therefore use synthetic self-supervision~\cite{choiNANSYUnifiedVoice2022,engel_self-supervised_nodate} or pitch-tracker labels~\cite{wu1DDSPbasedSingingVocoders2022,ycy2023golf-tismir}; other losses change STFT settings~\cite{MultiScaleSpectralLoss} or use optimal transport~\cite{torresUnsupervisedHarmonicParameter2024}, but their effect on synthesis is unclear.
Consequently, many systems rely on external $F_0$ estimators~\cite{ycy2024golf}, which are not differentiable, block joint training, and pass their errors into synthesis.

Most neural vocoders ignore phase because waveform misalignment is rarely audible~\cite{engelDDSPDifferentiableDigital2020}.
GOLF~\cite{ycy2023golf-tismir,ycy2024golf} is an exception: it synthesises with an all-pole filter and a glottal model, so its phase response follows the physics of the vocal tract.
This lets GOLF capture glottal pulses by waveform matching~\cite{ycy2023golf-tismir} to fine-tune the predicted parameters, enabling recovery of instantaneous phase and frequency.
GOLF itself cannot follow that path: its oscillator is still driven by an external $F_0$.

We introduce GOLF-inv with four contributions: (1) an additive glottal source that avoids wavetable resampling aliasing, with a closed-form pulse whose derivatives are analytic; (2) a phase correction from the Quasi-Harmonic Model (QHM)~\cite{pantazisPropertiesTimevaryingQuasiharmonic2008} that integrates frequency across frames without error accumulation; (3) a numerically stable state-space lattice filter evaluated by a generalised parallel prefix scan; and (4) unsupervised joint learning of $F_0$, phase, and envelopes from complementary spectral and harmonic-only waveform losses.
Throughout, reconstruction quality matters only as evidence that the inferred parameters explain the signal.
Audio examples are available online~\footnote{Available at \href{https://yoyolicoris.github.io/golf-inv-demo/}{yoyolicoris.github.io/golf-inv-demo/}}.

\section{Methodology}\label{sec:methodology}
\subsection{Model formulation}\label{ssec:model}
Building on GOLF~\cite{ycy2024golf}, the proposed synthesiser represents the voice $S(z)$ using a glottal source $G(z)$, a shaped noise component $Q(z)$, and a vocal-tract filter $H(z)$:
\begin{align}
  S(z) & = H(z)\left(G(z) + Q(z)\right).
  \label{eq:model}
\end{align}
Here, $Q(z)$ is modelled as white-noise excitation passed through learned spectral shaping (\secref{ssec:noise}).
A neural encoder (\secref{ssec:encoder}) estimates all source and filter parameters from the input audio, trained by minimising reconstruction losses through this synthesiser.
Unlike GOLF, which needs external $F_0$ and a resampled wavetable, GOLF-inv builds the source additively, predicts frequency and phase per frame and reads $F_0$ off the corrected phase (\secref{ssec:glottal}), and replaces the direct-form filter with a state-space lattice (\secref{ssec:vocal_tract}).
Because the harmonic branch $H(z)G(z)$ has a deterministic phase response, it can be supervised by waveform mean squared error (MSE), unlike the stochastic $Q(z)$ branch.

\subsection{Training criterion}\label{ssec:training_criterion}
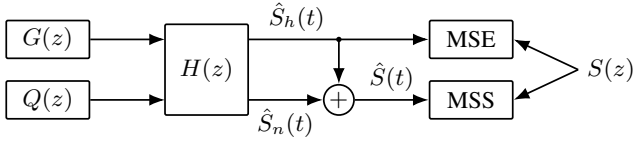
\begin{figure}[!ht]
  \centering
  \begin{tikzpicture}[
      >=Latex,
      x=1cm,
      y=1cm,
      line width=0.65pt,
      font=\small,
      block/.style={draw, rounded corners=0.6pt, minimum height=5mm, minimum width=11mm, align=center, inner sep=1.5pt},
      filter/.style={block, minimum height=12mm},
      loss/.style={block},
      sum/.style={draw, circle, minimum size=4mm, inner sep=0pt}
    ]
    \node[block] (harmonic) at (0.55,0.8) {$G(z)$};
    \node[block] (noise) at (0.55,0) {$Q(z)$};
    \node[filter] (filter) at (2.65,0.4) {$H(z)$};
    \coordinate (harmonic-out) at (3.2,0.8);
    \coordinate (noise-out) at (3.2,0);
    \coordinate (branch) at (4.4,0.8);
    \node[sum] (mix) at (4.4,0) {$+$};
    \node[loss] (mse) at (6.15,0.8) {MSE};
    \node[loss] (mss) at (6.15,0) {MSS};
    \node (target) at (8,0.4) {$S(z)$};

    \draw[->] (harmonic.east) -- (2.1,0.8);
    \draw[->] (noise.east) -- (2.1,0);
    \draw[->] (harmonic-out) -- node[pos=0.27, above] {$\hat{S}_h(t)$} (mse.west);
    \draw[->] (noise-out) -- node[midway, below] {$\hat{S}_n(t)$} (mix.west);
    \draw[->] (branch) -- (mix.north);
    \fill (branch) circle[radius=1pt];
    \draw[->] (mix.east) -- node[midway, above] {$\hat{S}(t)$} (mss.west);
    \draw[->] (target.west) -- (mse.east);
    \draw[->] (target.west) -- (mss.east);
  \end{tikzpicture}
  \caption{Loss routing in GOLF-inv. MSE compares only $\hat{S}_h(t)$ with $S(z)$, whereas MSS compares the spectrum of the full reconstruction $\hat{S}(z)$ with the target $S(z)$.}
  \label{fig:loss_path}
\end{figure}

The sum of the deterministic harmonic output $\hat{S}_h(z)=H(z)G(z)$ and a stochastic output $\hat{S}_n(z)=H(z)Q(z)$ forms the full reconstruction $\hat{S}(z)$.
These outputs define the two complementary loss paths in \figref{fig:loss_path}.
Multi-scale spectral loss (MSS)~\cite{MultiScaleSpectralLoss} is applied to the full reconstruction $\hat{S}(z)$, so both harmonic and stochastic magnitudes are learned.
Waveform MSE constrains phase, but the random excitation in $\hat{S}_n(z)$ makes its waveform unpredictable.
We therefore exclude $\hat{S}_n(z)$ from MSE and apply it only to the deterministic harmonic output $\hat{S}_h(z)$.
This works well in practice, likely because voiced regions dominate our training data.
The total objective sums harmonic-only waveform MSE and $\lambda$-weighted full-signal MSS.

\subsection{The glottal source \texorpdfstring{$G(z)$}{G(z)}}\label{ssec:glottal}
GOLF resamples a non-band-limited LF glottal-pulse wavetable~\cite{fantFourparameterModelGlottal,ycy2023golf-tismir}, which can introduce aliasing artefacts.
Later versions oversample~\cite{ycy2024golf} to mitigate it, but cannot remove it.
In this work, we generate the glottal source using additive synthesis:
\begin{align}
  \label{eq:glottal_source}
  g(t) & = \sum_{k=1}^K a_k \cos\left(\phi_k(t)\right).
\end{align}
We assume one canonical pulse shape, so the amplitudes $a_k$ and phase offsets $\psi_k$ are global and time-invariant.
$\phi_k(t)$ is computed from the instantaneous phase $\phi(t)$ and the offset:
\begin{align}
  \label{eq:phase_offset}
  \phi_k(t) & = k\phi(t) + \psi_k.
\end{align}
We set $\psi_1 = 0$ so the fundamental harmonic is in phase with $\phi(t)$.
$K$ is set to 100 in this work, and we band-limit the glottal source by setting $a_k = 0$ when $\frac{d\phi_k(t)}{dt} \ge \pi$, so no partial crosses Nyquist.
We constrain $\sum_{k=1}^K a_k^2 = 1$ and $a_1 > a_2 > \dots > a_K > 0$ so the first harmonic stays strongest and $\phi(t)$ tracks the fundamental rather than a sub-harmonic.

To obtain $\phi(t)$ without accumulating integration errors across frames, we adopt the phase correction mechanism from QHM~\cite{pantazisPropertiesTimevaryingQuasiharmonic2008,chenQHMGANNeuralVocoder2024}.
The encoder predicts frame-wise frequency $\hat{f}(t_l)$ and phase $\phi(t_l)$ at frame centres $t_l$.
$\hat{f}(t_l)$ is then upsampled to $\hat{f}(t)$ using cubic spline interpolation.
For $t \in [t_l,t_{l+1})$, we integrate the instantaneous frequency from the predicted phase at the current frame centre to obtain the uncorrected phase
\begin{align}
  \Tilde{\phi}(t) & = \phi(t_l) + \int_{t_l}^t 2\pi\hat{f}(\tau)\,d\tau.
  \label{eq:instantaneous_phase}
\end{align}
The final instantaneous phase $\phi(t)$ is obtained by smoothly distributing the boundary phase mismatch $z_l$:
\begin{subequations}
  \label{eq:phase_correction}
  \begin{gather}
    \phi(t) = \Tilde{\phi}(t) + \frac{z_l}{2}\left(1 - \cos\left(\frac{\pi(t - t_l)}{t_{l+1} - t_l}\right)\right) \\
    z_l = \left[\phi(t_{l+1}) - \Tilde{\phi}(t_{l+1})\right]_{2\pi},
  \end{gather}
\end{subequations}
where $[x]_{2\pi} = (x + \pi) \bmod 2\pi - \pi$ wraps phase differences to $[-\pi, \pi)$.

However, the instantaneous frequency $\frac{1}{2\pi}\frac{d\phi(t)}{dt}$ can deviate from $\hat{f}(t_l)$.
The maximum absolute frequency correction in \eqnref{eq:phase_correction} occurs at $t = \frac{t_l + t_{l+1}}{2}$ and equals $\frac{|z_l|}{4(t_{l+1} - t_l)}$.
With a \SI{100}{\hertz} frame rate, that roughly means $\pm$\SI{79}{\hertz} maximum correction, significant for low-pitched voices.
The encoder can thus leverage both $\hat{f}(t_l)$ and $\phi(t_l)$ to encode $F_0$.
To decouple the two, we randomly adjust the integration window to be $[t_l, t_{l+m}) $, where $ m \sim \mathcal{U}\{1,2,3,4\}$ during training as a regulariser.
In this way, the maximum correction is divided by $m$, and the randomness forces the encoder to rely more on $\hat{f}(t_l)$ for phase conversion.
We bound $\hat{f}(t_l)$ within \SIrange[range-units=single]{32}{960}{\hertz}.

\subsection{The noise source \texorpdfstring{$Q(z)$}{Q(z)}}
\label{ssec:noise}
Following previous design~\cite{ycy2024golf}, $Q(z)$ denotes white noise after learned spectral shaping.
We parametrise the shaping response using cepstra, similar to Liu et al.~\cite{liuNeuralHomomorphicVocoder2020}, but with symmetric real coefficients, yielding zero phase.
We use 30 cepstral coefficients, sufficient for the broad shape of the noise spectrum, leaving fine detail to $H(z)$.
The frame-wise frequency response is converted to a time-domain FIR via the windowing method, using an FFT size of 512 and a 149-point Hamming window.

\subsection{The vocal tract filter \texorpdfstring{$H(z)$}{H(z)}}\label{ssec:vocal_tract}
The all-pole vocal tract filter models the voice's time-varying dynamics.
Direct-form implementations~\cite{ycy2024golf} accumulate numerical error at high-dynamic-range summation nodes~\cite{jacksonInteractionRoundoffNoise1970}.
The vocal tract can be discretised as a cascade of cylindrical acoustic tubes, with each area discontinuity scattering waves into forward and backward components~\cite{markelLinearPredictionSpeech1976}.
A lattice section models the scattering at one junction, so its reflection coefficient follows from the adjacent areas, $\kappa_n(t) = \frac{A_{n-1}(t) - A_n(t)}{A_{n-1}(t) + A_n(t)}$.
Since $A_n(t)>0$, $\kappa_n(t)$ stays in $(-1,1)$, tying the parametrisation to vocal-tract geometry and bounding the section weights, so the summation nodes avoid the dynamic range that troubles the direct form.

\subsubsection{State-space realisation of the lattice filter}
We use a state-space lattice realisation~\cite{morfStatespaceStructuresLadder1980}, which enables parallel evaluation.
With $\bx(t) = [x_N(t), x_{N-1}(t), \dots, x_1(t)]^\top$ denoting the state vector, an $N^\textit{th}$ order lattice filter can be expressed as follows:
\begin{subequations}
  \label{eq:ssm}
  \begin{align}
    \bx(t+1) & = \bA(t)\bx(t) + \bB(t)e_0(t) \label{eq:ssm_recursion} \\
    y(t)     & = \bC^\top(t)\bx(t) + e_0(t)
  \end{align}
  \begin{gather}
    \bA       =
    \begin{bmatrix}
      \kappa_N   & \kappa_{N-1}          & \cdots & \kappa_2              & \kappa_1              \\
      \kappa_N^c & -\kappa_{N-1}\kappa_N & \cdots & -\kappa_2\kappa_N     & -\kappa_1\kappa_N     \\
      0          & \kappa_{N-1}^c        & \cdots & -\kappa_2\kappa_{N-1} & -\kappa_1\kappa_{N-1} \\
      \vdots     & \vdots                & \ddots & \vdots                & \vdots                \\
      0          & 0                     & \cdots & \kappa_{2}^c          & -\kappa_1\kappa_2
    \end{bmatrix} \\
    \bB  =
    \begin{bmatrix}
      1 & -\kappa_N & -\kappa_{N-1} & \cdots & -\kappa_2
    \end{bmatrix}^\top
  \end{gather}
\end{subequations}
where $\kappa_n^c = 1 - \kappa_n^2$ and we have omitted $t$ for brevity.
$N=14$ roughly corresponds to the length of a male vocal tract at \SI{16}{\kilo\hertz}~\cite{sudholtVocalTractArea2023}.
We learn $\bC(t) \in \RR^N$ to incorporate zeros and apply a time-varying pre-filter gain $\gamma(t)$ before the filter; the direct-path gain is fixed to 1.

\subsubsection{Parallel implementation}\label{ssec:parallel_scan}
Evaluating \eqnref{eq:ssm_recursion} directly takes $T$ sequential steps.
Composing $L$ consecutive steps into a single affine transform advances the state every $L$ steps:
\begin{subequations}
  \label{eq:reduced_recursion}
  \begin{align}
    \Tilde{\bA}_i(t) & = \prod_{j=i}^{L-1}\bA(t+j), \quad \Tilde{\bA}_L(t)=\bI \label{eq:reduced_A}  \\
    \bu(t)           & = \sum_{i=0}^{L-1} \Tilde{\bA}_{i+1}(t)\bB(t+i)e_0(t+i) \label{eq:reduced_bu} \\
    \bx(t+L)         & = \Tilde{\bA}_0(t)\bx(t) + \bu(t). \label{eq:block_transition}
  \end{align}
\end{subequations}
Products are time-ordered, with factors of larger $j$ placed to the left.
The block coefficients in \eqnref{eq:reduced_A} and \eqnref{eq:reduced_bu} depend only on their own block, so all blocks are computed in parallel, leaving in \eqnref{eq:block_transition} a sequential chain $L$ times shorter.
Once the block-boundary states $\bx(nL)\mid_{n \in \mathbb{N}_+}$ are known, \eqnref{eq:ssm_recursion} recovers the remaining states in parallel across blocks, with $L-1$ sequential steps per block.
The reduced recursion has the same affine form as \eqnref{eq:ssm_recursion}, so the same reduction applies to itself; recursing until one step remains lowers the sequential cost from $O(T)$ to $O(L\log_L T)$.
For $L=2$, this is the standard parallel prefix scan~\cite{BlellochTR90}; we use $L=32$, selected empirically for runtime.
Our pure-PyTorch implementation is sufficiently fast for this work~\footnote{Available at \href{https://github.com/yoyolicoris/philtorch}{github.com/yoyolicoris/philtorch}.}.

\subsection{The analysis encoder}\label{ssec:encoder}
The encoder follows the same architecture as in~\cite{ycy2024golf}, but with \SI{16}{\kilo\hertz} audio in this work.
The input is a 4-channel spectrogram (FFT size 512, \SI{100}{\hertz} frame rate) holding the log-magnitude, angle, real and imaginary parts.
This STFT feature is used only by the analysis encoder to extract synthesis parameters; the decoder remains the physical source-filter synthesiser in \eqnref{eq:model}.
The four channels are redundant, which is acceptable for an analysis front end, unlike DDSP, which uses a mel-spectrogram to condition synthesis~\cite{engelDDSPDifferentiableDigital2020,wu1DDSPbasedSingingVocoders2022}.
The feature passes through four 2D convolutional layers, each followed by batch normalisation, ReLU activation, and max pooling along the frequency axis.
Then, the frequency axis and channels are flattened and fed into three Bi-LSTM layers with 256 hidden units and a dropout rate of 0.1.
After layer normalisation, fully connected layers project to the parameters, with non-linearities keeping them in range.
The parameters are linearly upsampled to the audio rate for time-varying synthesis.

\section{Experiments}
\subsection{Datasets and training details}
We evaluate the proposed system on two singing voice datasets, M4Singer~\cite{zhangM4SingerMultiStyleMultiSinger} and LM-SSD~\cite{schwarDatasetLarynxMicrophone2024}, both resampled to \SI{16}{\kilo\hertz}.
M4Singer contains \SI{30}{\hour} of Chinese singing voice from 20 singers, ranging from Bass to Soprano.
For each singer, \SI{10}{\percent} of the data is randomly assigned to the test set, and the remainder is used for training; the split is therefore within-singer.
LM-SSD contains singing performances recorded simultaneously with a larynx microphone and a studio microphone.
We select only solo recordings made with a studio microphone and without cross-talk, yielding three singers and roughly two hours of data.
The larynx microphone captures throat vibration directly with little cross-talk and provides the reference labels (\secref{ssec:f0_estimation}, \secref{ssec:gci_detection}); all methods process the studio signal.
For each song, we hold out the sixth take for testing, and the remaining takes form the training set; thus, the test split contains unseen performances, but not unseen songs or singers.

Audio is peak-normalised and cut into 32768-sample segments; we train with AdamW, learning rate 0.0002, batch size 8.
Following~\cite{MultiScaleSpectralLoss}, the spectral loss uses co-prime window and hop sizes, \{251, 509, 1021, 2053\} and \{67, 127, 257, 521\}, which smooths the loss surface and helps $F_0$ learning.
We compute it with \texttt{auraloss} and set $\lambda=0.01$ from early experiments.
The random integration window (\secref{ssec:glottal}) hurts early convergence, so we train 10 epochs without it, then switch it on until validation loss stops improving.

\subsection{Model configurations}
\textbf{GOLF-inv} is the proposed unsupervised system.
\textbf{GOLF} uses $F_0$ from FCNF0++~\cite{morrisonCrossdomainNeuralPitch2023} for the oscillator, as in the original vocoder~\cite{ycy2024golf}.
\textbf{DDSP+$\bf F_0$} follows the DDSP-inv setting of~\cite{engel_self-supervised_nodate}: FCNF0++ provides $F_0$ targets for the prediction head to follow, and time-varying additive synthesis replaces $H(z)$.
\textbf{QHM} is our conventional analysis and resynthesis reference and we use FCNF0++'s $F_0$ as initialisation for its analysis process~\cite{pantazisPropertiesTimevaryingQuasiharmonic2008}.
Additional ablations replace the lattice with the direct-form filter of~\cite{ycy2024golf}, omit $H(z)$ to use a QHM-only source, or disable integration-window randomisation.

\subsection{Copy-synthesis evaluation}\label{ssec:copy_synth}
\begin{figure}[!t]
  \centering
  \includegraphics[width=\columnwidth]{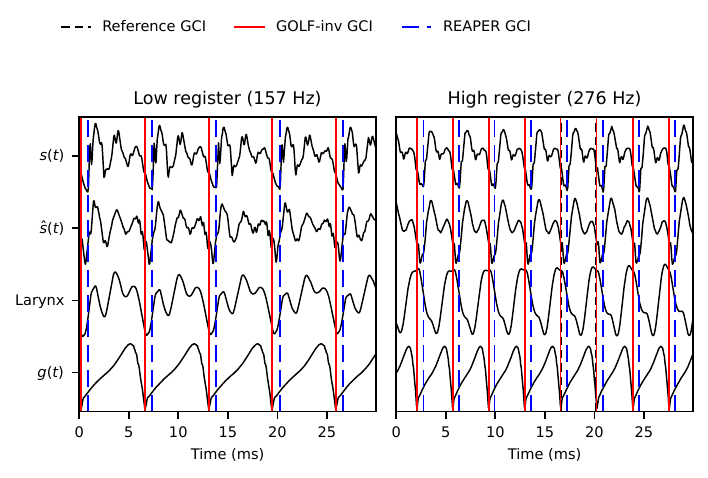}
  \caption{Synchronised waveforms and GCIs on two test LM-SSD excerpts. The reconstruction $\hat{s}(t)$ follows the target $s(t)$ cycle by cycle.}
  \label{fig:glottal_gci}
\end{figure}

Copy-synthesis reconstructs each recording from parameters inferred from that same recording through the analysis encoder and physical synthesiser.
We report MSS, mel-cepstral distortion (MCD), perceptual evaluation of speech quality (PESQ), and $\text{SRER}=10\log_{10}\frac{\sum_t s^2(t)}{\sum_t(s(t)-\hat{s}(t))^2}$.
The first three measure spectral fidelity; SRER measures sample-level waveform match, and thus phase. The two need not move together.
A positive SRER requires the reconstruction to follow the target sample by sample, which phase-agnostic systems cannot do; \figref{fig:glottal_gci} shows $\hat{s}(t)$ matches $s(t)$ closely on a low- and a high-register LM-SSD excerpt.
We compute each metric per clip and average over the split, weighted by clip duration.
In \tabref{tab:reconstruction_quality}, GOLF gives the lowest spectral distortion using external $F_0$, and QHM the highest SRER, but QHM fits each frame by least squares from an external $F_0$ and has no vocal tract.
Its SRER lead comes mostly from the noise model: QHM fits sinusoids to unvoiced parts too, while our noise branch $Q(z)$ has random phase and leaves an irreducible residual there.
GOLF-inv outperforms DDSP+$F_0$ on every metric and changes SRER from \SI{-2.67}{\dB} to \SI{8.10}{\dB} despite using no $F_0$ targets.
Filter ablations yield \SI{7.92}{\dB} SRER with the direct form.
The QHM-only source reaches \SI{10.68}{\dB} SRER but worsens MCD to \SI{6.21}{\dB}, so the source-filter split is needed for synthesis quality.
\begin{table}[!ht]
  \centering
  \small
  \caption{Synthesis evaluation on the test split of M4Singer and LM-SSD datasets.}
  \label{tab:reconstruction_quality}
  \begin{tabular}{lrrrr}
    \toprule
    Model      & MSS($\downarrow$) & MCD($\downarrow$) & PESQ($\uparrow$) & SRER($\uparrow$)   \\
    \midrule
    GOLF-inv   & 0.981             & 4.469 dB          & 2.613            & 8.103 dB           \\
    \midrule
    GOLF       & \textbf{0.868}    & \textbf{3.917} dB & 2.838            & -2.801 dB          \\
    DDSP+$F_0$ & 1.072             & 5.175 dB          & 2.485            & -2.665 dB          \\
    QHM        & 0.871             & 4.668 dB          & \textbf{2.967}   & \textbf{13.745} dB \\
    \bottomrule
  \end{tabular}
\end{table}

\subsection{Fundamental frequency estimation}\label{ssec:f0_estimation}
We evaluate $F_0$ estimation on LM-SSD against the larynx reference, where we use a majority vote of REAPER~\cite{google_reaper}, pYIN~\cite{mauchPYINFundamentalFrequency2014}, and SWIPE~\cite{camacho2008sawtooth} estimates.
For GOLF-inv, $F_0$ is the frame average of $\frac{1}{2\pi}\frac{d\phi(t)}{dt}$ under a local-stationarity assumption.
A frame is marked as voiced if $\|\hat{S}_h(z)\|^2 > \|\hat{S}_n(z)\|^2$ and the energy is above \SI{-40}{\dB}.
We report raw pitch accuracy (RPA), the share of frames within 50 cents of the target, and overall accuracy (OA), which also counts unvoiced frames.
Frame counts are pooled over the split before each percentage is formed, rather than averaging per-clip scores.
Baselines are SWIPE, pYIN, REAPER, and two neural pitch trackers, FCNF0++~\cite{morrisonCrossdomainNeuralPitch2023} and PESTO~\cite{riouPESTOPitchEstimation2023}, using their published model weights.
We report the training split as an in-sample diagnostic: GOLF-inv trains on LM-SSD while the baselines run off the shelf, so part of its lead comes from adapting to this data, which needs no $F_0$ labels and could therefore be repeated on any corpus.

In \tabref{tab:f0_estimation}, GOLF-inv outperforms all neural baselines (DDSP+$F_0$, FCNF0++, and PESTO) and achieves the highest overall accuracy on both splits, with RPA exceeding REAPER and SWIPE.
pYIN attains the highest RPA but the lowest OA, so its accuracy on voiced frames comes with weaker voicing decisions.
Without the random window, test RPA from $\hat{f}(t_l)$ drops 5.39 points to \SI{86.89}{\percent}, against \SI{92.28}{\percent} with correction; with the random window, the same gap is only 0.50 points, demonstrating that the randomisation prevents the phase correction from absorbing frequency error.
\begin{table}[!ht]
  \centering
  \small
  \caption{$F_0$ estimation results (\%) on the LM-SSD dataset.}
  \label{tab:f0_estimation}
  \begin{tabular}{lcccc}
    \toprule
    \multirow{2}{*}{Model}                              & \multicolumn{2}{c}{Train} & \multicolumn{2}{c}{Test}                                          \\
    \cmidrule(lr){2-3} \cmidrule(lr){4-5}
                                                        & RPA    ($\uparrow$)       & OA    ($\uparrow$)       & RPA   ($\uparrow$) & OA   ($\uparrow$) \\
    \midrule
    GOLF-inv                                            & 92.45                     & \textbf{95.19}           & 92.91              & \textbf{95.79}    \\
    \hspace{0.5em}$\hookrightarrow$~$\hat{f}(t_l)$ only & 91.79                     & 94.85                    & 92.41              & 95.54             \\
    \midrule
    Fixed window                                        & 91.82                     & 94.83                    & 92.28              & 95.44             \\
    \hspace{0.5em}$\hookrightarrow$~$\hat{f}(t_l)$ only & 86.43                     & 92.07                    & 86.89              & 92.75             \\
    \midrule
    DDSP+$F_0$                                          & 91.19                     & 93.29                    & 91.52              & 93.96             \\
    FCNF0++                                             & 90.85                     & 92.72                    & 90.78              & 93.64             \\
    PESTO                                               & 92.01                     & 95.04                    & 92.33              & 95.52             \\
    \midrule
    pYIN                                                & \textbf{94.91}            & 90.00                    & \textbf{94.91}     & 91.63             \\
    SWIPE                                               & 91.27                     & 91.69                    & 92.06              & 93.85             \\
    REAPER                                              & 92.05                     & 94.04                    & 92.61              & 95.63             \\
    \bottomrule
  \end{tabular}
\end{table}
\subsection{Glottal closure instant detection}\label{ssec:gci_detection}
Finally, we evaluate glottal closure instant (GCI) detection on the LM-SSD test set against REAPER pitch marks computed on the larynx signal.
Because \eqnref{eq:glottal_source} is a closed-form sum of sinusoids, its first and second derivatives are analytic, so the Newton-Raphson method locates the minimum of the learned canonical pulse, which we take as its closure phase; the corresponding instants in $\phi(t)$ are marked as GCIs.
\figref{fig:glottal_gci} also plots the larynx signal and the learned source $g(t)$, with the estimated and reference GCIs overlaid.
Following~\cite{naylor_estimation_2007}, the identification rate (IDR), false alarm rate (FAR), missed rate (MR), and the standard deviation of the timing error ($\sigma$) are reported in \tabref{tab:gci_estimation}.
A mark counts as correctly identified when it falls inside a one-period window centred on the reference GCI.
Hits, false alarms, and misses are likewise pooled over the split, and $\sigma$ is taken over the pooled timing errors.
Different modelling assumptions can displace estimates from the reference by a near-constant offset, as REAPER's marks show in \figref{fig:glottal_gci}.
As a standard deviation, $\sigma$ ignores such an offset and measures the spread that remains.
GOLF-inv tracks the reference in both registers, trailing REAPER by 0.53 points in IDR and \SI{0.01}{\milli\second} in $\sigma$.
\begin{table}[!ht]
  \centering
  \small
  \caption{GCI detection performance on LM-SSD test set.}
  \label{tab:gci_estimation}
  \begin{tabular}{lcccc}
    \toprule
    Model    & IDR ($\uparrow$) & FAR ($\downarrow$) & MR ($\downarrow$) & $\sigma$ ($\downarrow$) \\
    \midrule
    GOLF-inv & 96.93\%          & 1.10\%             & 1.97\%            & 0.77 ms                 \\
    \midrule
    REAPER   & \textbf{97.46}\% & \textbf{0.82}\%    & \textbf{1.72}\%   & \textbf{0.76} ms        \\
    \bottomrule
  \end{tabular}
\end{table}

\subsection{Discussion}
The experiments show that a strong inductive bias in the synthesiser, plus the waveform loss, is enough to learn instantaneous phase and frequency without labels.
Convergence depends on initialising the predicted frequency range to overlap the data and on harmonic amplitudes decaying with frequency.
The lattice gives a higher SRER than the direct form, which also needs gradient clipping to converge.
The QHM-only source is less stable to train, supporting the source-filter decomposition.
Each output sample of GOLF-inv sums up to 100 sinusoids where a wavetable lookup would read a single value; in exchange, the source avoids aliasing, and its closed form allows analysis, such as the Newton search of \secref{ssec:gci_detection}.
The main limitation is the biased criterion: applying MSE only to the deterministic path encourages harmonic energy where noise should dominate, audible as residual buzzing in unvoiced regions, which voicing-dependent masking or spectral whitening before the MSE may reduce.
We only test singing voice; many instruments share a similar source-filter structure, and speech is left for future work.

\section{Conclusion}
Fitting a physically constrained synthesiser and neural encoder to a waveform is enough to recover the parameters of voice production: the $F_0$ and closure instants here are read off the encoder, and on LM-SSD they match REAPER with no $F_0$ label.
The proposed method works because it matches the waveform, including phase, rather than the magnitude spectrum alone.
Within the limits discussed above, inverse synthesis offers a practical route to label-free voice analysis.

\section{ACKNOWLEDGMENT}
Yu is a research student at the UKRI CDT in AI and Music, jointly supported by UKRI (grant EP/S022694/1) and QMUL.
Fazekas was supported by the Leverhulme Trust and the Royal Academy of Engineering under the RAEng / Leverhulme Trust Research Fellowships scheme.
Claude Sonnet 5 was used to create \figref{fig:glottal_gci}.
Claude Opus 5 and Copilot with Gemini 3.7/3.8 were used to trim the initial draft.

\section{COMPLIANCE WITH ETHICAL STANDARDS}
This is a non-commercial secondary analysis of publicly released recordings, M4Singer (CC BY-NC-SA 4.0) and LM-SSD (CC BY 4.0), for which no ethical approval was required.

\end{document}